# Nonlinear bicolor holography using plasmonic metasurfaces

*Daniel Frese[1], Qunshuo Wei[2], Yongtian Wang[2],*
*Mirko Cinchetti[3], Lingling Huang[2], and Thomas Zentgraf[1]*

[1] Department of Physics, Paderborn University, Warburger Straße 100, 33098 Paderborn, Germany

[2] School of Optics and Photonics, Beijing Institute of Technology, 100081, Beijing, China

[3] Experimentelle Physik VI, Technische Universität Dortmund, 44227 Dortmund, Germany

**Abstract:** Nonlinear metasurface holography shows the great potential of metasurfaces to control the phase, amplitude, and polarization of light while simultaneously converting the frequency of the light. The possibility of tailoring the scattering properties of a coherent beam, as well as the scattering properties of nonlinear signals originating from the meta-atoms facilitates a huge degree of freedom in beam shaping application. Recently, several approaches showed that virtual objects or any kind of optical information can be generated at a wavelength different from the laser input beam. Here, we demonstrate a single-layer nonlinear geometric-phase metasurface made of plasmonic nanostructures for a simultaneous second and third harmonic generation. Different from previous works, we demonstrate a two-color hologram with dissimilar types of nanostructures that generate the color information by different nonlinear optical processes. The amplitude ratio of both harmonic signals can be adapted depending on the nanostructures' resonance as well as the power and the wavelength of the incident laser beam. The two-color holographic image is reconstructed in the Fourier space at visible wavelengths with equal amplitudes using a single near-infrared wavelength. Nonlinear holography using multiple nonlinear processes simultaneously provides an alternative path to holographic color display applications, enhanced optical encryption schemes, and multiplexed optical data-storage.

**Introduction**

Metasurface holography became a key technology for many light shaping applications. With the invention of computer-generated holograms (CGHs) in 1966 by Brown and Lohmann [1], the elaborate optical recording process of conventional holograms were extended by computational algorithms. Thus, virtual objects can now be reconstructed by a hologram using for example spatial-light modulators (SLMs) [2-3] or metasurfaces made of nanostructures [4-5]. Compared to SLMs, metasurfaces have several advantages: the information density is higher due to the subwavelength dimension of the structured meta-atoms that can be compared with a pixel of the SLM, followed by the elimination of higher diffraction orders and high image quality [6]. Furthermore, the subwavelength scale of the devices open-up new possibilities in integrated on-chip applications through planar fabrication technics. Alongside phase and amplitude holograms [7-8], the holographic method spread out in many applications like holographic displays technologies [9], metasurface imaging [10-11], information processing with particular security features using chiral structures [12-13], polarization optics [14-15], mode shaping [16-17], and quantum optics [18-20]. Recently, systems based on transition metal dichalcogenide (TDMC) have demonstrated that beam shaping and information processing can be realized at dimensions of atomic thickness and hybrid structures [21-23].

A pursued challenge is the realization of colored holograms for holographic displays. Therefore, several techniques like the subdivision of pixels in RGB components [24-25], angular dependent color holography [26], and different sized silicon blocks multiplexed for the phase modulation of different wavelengths or color printing [27] are only a few examples. These holograms require different laser sources (for instance red, green and blue) for the reconstruction of colored images. However, another possibility is to utilize nonlinear processes of particular nanostructures together with a single wavelength laser beam. For instance, based on harmonic generation processes of meta-atoms, the frequency of the incident laser beam can be converted to higher orders and its response can be phase modulated [28-30]. Recently, nonlinear holograms based on a single nonlinear process like second harmonic generation (SHG) or third harmonic generation (THG) already show the capability in nonlinear wavefront shaping for complex image formation [29, 31-33]. Combined with polarization multiplexing technologies in holographic applications, the information density per unit area can be enhanced and different images can be stored in a single 2D device [32, 34].

Here, we present a nonlinear metasurface hologram in the visible regime, utilizing SHG and THG simultaneously. Our holographic phase-only reconstruction scheme is based on the nonlinear

Pancharatnam-Berry-phase introduced by plasmonic nanostructures. The used two- and three-fold rotation symmetric nanostructures (C2 and C3) are oriented according to the desired local phase for the second- and third-harmonic light in the cross-circular polarization state. Thus, the appearance of unwanted image fragments in the co-polarization channel can be avoided by additional polarization filtering. The hologram can be reconstructed using a single-wavelength near-infrared laser beam. An important part of the metasurface design is that the nonlinear responses of the second and third-order have in general different strength and dependence of the incident electric light field. Therefore, we show that the design can be adapted to enable nonlinear hologram reconstruction with equal amplitude distributions for second and third harmonic generation simultaneously. By choosing the same number of C2 and C3 antennas, the highest information density per area is achieved. Our approach offers an alternative way to conventional color holography.

**Metasurface design and working principle**

With our work, we demonstrate a bicolor nonlinear metasurface hologram with equal intensities of phase-modulated second and third harmonic sources, as illustrated in Figure 1. The metasurface, consisting of periodically arranged gold nanostructures, carries the encoded images of a tree and a house with different colors. The images are reconstructed in the visible spectral range when the meta-hologram is illuminated with near-infrared light. The image colors red and blue correspond to the second and third-order process, respectively, arising from the two different meta-atom types. For full phase range coverage of the converted light in the desired polarization channel, we designed and fabricated a metasurface based on the geometric Pancharatnam-Berry phase. When illuminated with circularly polarized light, spatially varying meta-atom rotations can introduce a 0 to $2\pi$ phase shift in the linear and nonlinear regime, depending on the meta-atom rotation symmetry and its orientation angle $\theta$ with respect to the lab frame. For the desired harmonic generation processes, we consider the selection rules for harmonic processes based on the macroscopic nonlinear material polarization, which holds in a good approximation for plasmonic nanostructures [35]. For instance, a single m-fold rotation symmetric meta-atom supports harmonic generations of the order $n = l(m \pm 1)$, where $n$ is the harmonic order, $l$ an arbitrary integer and $m$ the rotation symmetry of the meta-atom. Such type of meta-atoms belongs to the cyclic group and are conventionally named C$m$ for $m$-fold rotation symmetry. To realize a nonlinear colored hologram in the circular cross-polarization channel, we select the C2 and the C3 symmetry. The optical excitation of these nanostructures and the fundamental principle of continuous phase control of nonlinear signals originating from them are theoretically described and experimentally verified previously [30, 35]. In general, the introduced cross-polarized SHG phase of a C3 meta-atom rotated by the angle of $\theta$ is $\varphi_{SHG,C3} = -\sigma 3\theta$, while for THG of a C2 meta-atom it is $\varphi_{THG,C2} = -\sigma 4\theta$ [30-31]. The parameter $\sigma = \pm 1$ indicates the handiness of the incident laser beam (left or right circularly polarized), respectively. Utilizing these two types of meta-atoms within a single metasurface, a two-colored hologram based on second and third harmonic generation is realized.

An important factor in designing a nonlinear colored hologram based on different harmonic processes is the strength of the different harmonic signals. If both nonlinear processes can be tuned to the same order of magnitude, applications can benefit from harmonic multiplexing technologies, for example in nonlinear holography and optical information processing. For a nonlinear color-hologram, we desire for example equal amplitudes for both colors. By consideration of the power series of the meta-atoms material polarization [36], one can find that each nonlinear process of the order $n$ depends on the susceptibility $\chi_n(\omega)$ and scales with the electric field $E(\omega)$ of the incident laser light with the $n$-th power. Thus, to generate a balanced nonlinear multi-harmonic metasurface hologram based on different harmonic generations, generally, three parameters can be adjusted: the power of the incident light field, the resonance frequency $\omega_{res}$ of the different involved plasmonic meta-atoms (which influences $\chi_n(\omega)$), and the filling ratio of the different meta-atom types. A passive way to influence the imaged signal strength is the usage of color filters to attenuate signal components to the desired value. In our case, we actively tune the three parameters, so that $\chi_2^{C3}(\omega)E^2(\omega) \approx \chi_3^{C2}(\omega) E^3(\omega)$, while $\chi_2^{C3}(\omega)$ is the susceptibility of the C3 meta-atoms, $\chi_3^{C2}(\omega)$ the susceptibility of the C2 meta-atoms.

Furthermore, to encode a two-colored Fourier hologram into the two-antenna-type metasurface, we use the modified parallel iterative Gerchberg-Saxton algorithm, which can obtain an optimized phase-only hologram with arbitrary pixel arrangements [27]. In our design, the same number of C2 and C3 meta-atoms are allocated arbitrarily on the metasurface whereas their orientation angles $\theta$ with respect to the lab frame record the phase information $\varphi_{SHG,C3}(\theta)$ and $\varphi_{THG,C2}(\theta)$ of each hologram. Note, that in our case of a square lattice of meta-atoms, we end up with a square array of spatially distributed rotation angles. According to the nonlinear Pancharatnam Berry-phase, the rotation angles of the C2 and C3 antennas enable a continuous $2\pi$ phase coverage for both nonlinear signals. Parallel iterative loops between the two Fourier holograms and the target images of two different harmonic signals are constructed independently via the Fourier transform propagating function. Considering the chosen arrangements of different structures, our algorithm only retains the phase distributions within the irregular shapes while setting the amplitude and phase to be zero at the other regions. Because each sub hologram is optimized to the shape of the corresponding arrangements, different target images can be reconstructed with good wavelength selectivity and high quality while the crosstalk can be avoided to the largest extent.

The localized plasmon-polariton resonance of both structures is tailored by the antenna geometries. Therefore, we simulated the transmittance of different C2 and C3 structures with periodic boundary conditions using the finite-difference time-domain (FDTD) method, to find structure dimensions, which result in a localized plasmon-polariton resonance at about 1300 nm (Fig. 3A, dashed lines). For the nanoantenna material properties, we choose the optical constants of gold by Johnson and Christy [37] and the refractive index $n = 1.4$ for the substrate. In general, the period of the metasurface array should be smaller than the wavelengths involved, to avoid diffraction effects. However, the minimum unit cell size is limited by the dimensions of the nanoantennas themselves, as well as a minimum distance to avoid coupling between the nearest neighbors. As a compromise, we chose a 500 nm period to avoid diffraction of the fundamental and SHG wavelength. For the THG, the image should also appear in the first diffraction order, which is outside the NA of our microscope objective.

Experimentally, the linear transmission of the fabricated metasurface is measured with a Fourier-transform infrared spectrometer. The C2 only (blue) and C3 only (red) structures have been investigated separately to check the plasmon-polariton resonance of the fabricated structure (Fig. 2A). Compared to the simulation, the spectral position of the fabricated structures' resonance is slightly blue-shifted due to fabrication tolerances. However, the desired overlap of the C2 structures' resonance and the C3 structures' resonance is determined. After the combination of the two kinds of structures into a single metasurface, a single resonance dip (black) appears with a minimum at about 1260 nm and a transmission amplitude between the blue and red curves (Fig. 2A). In the ideal case, the transmission of the hologram is equal to the arithmetic mean of the C2 only and C3 only fields, if the hologram consists of an equal amount of both meta-atoms. Slight deviations can arise from fabrication tolerances and weak coupling between the meta-atoms.

The design parameters and the linear transmission of the metasurface are shown in Figure 2B. The fabricated metasurface consists of gold nanoantennas placed on a glass substrate. The fabrication is based on e-beam lithography. First, the meta-structure is written on a photoresist using an electron beam. After development, a 2-nm-thick chromium layer as an adhesion layer is deposited, followed by 30 nm of gold. After a lift-off process, only the written areas remain and form the metasurface. The fabricated C2 meta-atoms have an arm-length of 310 nm and the C3 meta-atoms of 170 nm. The hologram consists of 400 by 400 pixels with a unit cell size is 500 nm times 500 nm (Fig. 2B). Thus, the pixel density is $\frac{1}{2} \cdot \frac{1}{500 \text{ nm}} = 25400$ ppi for C2 and C3 meta-atoms. A scanning electron microscopy (SEM) image with the randomly distributed C2 and C3 meta-atoms is shown in Figure 2C. The orientation of the meta-atoms is crucial for the stored phase information within the metasurface design, while the distribution of the meta-atoms within the lattice can be arbitrarily arranged based on the above-mentioned modified Gerchberg-Saxton algorithm.

**Experimental results**

The following experiments are distinguished between two optical measurement systems: the imaging system and a spectrometer. The imaging system is used to reconstruct and measure the nonlinear Fourier-hologram by a camera, while the spectrograph is used to confirm the wavelength of the nonlinear signals. Since both systems are chromatic and have different spectral response functions and optical losses, we focus on the imaging system that visualizes the reconstructed hologram.

*Reconstruction of the nonlinear hologram*

For the optical reconstruction of the metasurface hologram, we use a Titanium-Sapphire pumped optical parametric oscillator as a tunable infrared light source. The pulsed laser light (pulse length 220 fs, repetition rate 80 MHz) first passes through a linear polarizer (LP) and a quarter-wave plate ($\lambda/4$) to define the desired circular input polarization state (Fig. 3A). The light is focused on the metasurface using a lens ($L_1$) with a focal length of $f = 200$ mm. The transmitted light is collected by a microscope objective (50x/0.42 Mitutoyo Plan Apo NIR). A heat absorption filter F1 blocks the fundamental wavelength to avoid damages to the imaging system by the high laser power. Two lenses ($L_{2,3}$) with a focal length of $f = 100$ mm in 4f arrangement are used to project the Fourier-plane to the camera. A second polarizer unit consisting of a quarter-wave plate followed by a linear polarizer is used as a polarization filter for transmitting only the circularly cross-polarized light. As a camera, we use a monochrome CMOS camera, due to the low RMS readout noise of 0.9 electrons per second, which is beneficial for nonlinear measurements which are usually accompanied by weak signal strengths. At position F2, different color filters can be placed to image the SHG and the THG signals separately. If no bandpass filter is used, the SHG and THG signals are imaged simultaneously in grayscale. Figure 3B shows the calculated resulting Fourier hologram image composed of both wavelengths. For better visualization, we used red for the SHG parts and blue for the THG parts of the image.

First, we reconstruct the nonlinear hologram over a wavelength range from 1240 nm to 1330 nm. The imaging range is limited by the camera's efficiency and the transmission of the optical components of the setup. Close to 400 nm, the detector's efficiency drops rapidly, as well as the transmission of the optical components. The upper limit is mainly limited by the near-infrared short pass filter, which protects the camera from the fundamental beam. Furthermore, the conversion efficiency of the plasmonic meta-atoms reduces when moving away from the resonance frequency. For probing the wavelength dependency, the laser peak power is set to 10.7 kW and the cameras' exposure time is set to 60 s. Figure 3C illustrates the SHG and THG conversion efficiency of the nonlinear hologram in dependence on the pump wavelength from 1240 to 1330 nm. The images show that the SHG parts of the image are strongest for 1240 nm and vanish for a longer wavelength (1330 nm, red dashed line), while the THG signal (blue dashed line) stays relatively constant in this wavelength range. This hints that the C2 and C3 meta-atoms show a different spectral nonlinear response, depending on the incident wavelength. We find that the SHG/THG ratio seems balanced at a wavelength of 1290 nm for the pump power of 10.7 kW, which is about 30 nm red-shifted compared to the combined holograms resonance dip shown in Figure 2A.

The reconstructed hologram for a pump wavelength of 1290 nm at 10.7 kW peak power for RCP illumination is shown in Fig. 3D. The exposure time is 270 s. If no bandpass color filter is used in front of the camera, the whole image is visible, confirming the equal signal strength of both nonlinear processes. By choosing a long-pass filter with a band edge of 550 nm, we confirm the spatial color composition that is illustrated in the simulated image (Fig. 3B). One can see, that only the SHG image appears, showing the treetop and house ground floor, as expected. In contrast, the replacement of the long-pass filter by a short-pass filter with a band edge of 500 nm, results only in the THG image. Note that for better visibility the contrast of the THG signal image is enhanced since the short-pass filter has a lower transmittance than the long-pass filter. Apparently, the different wavelengths in the image originate from the C3 and C2 meta-atoms, as they carry the correct phase information for reconstructing the different parts of the image. Hence, crosstalk originating from the combined hologram structure is not apparent. SHG originating from the C2 structures and THG originating from the C3 structures can appear due to unideal fabricated structures and surface roughness of the nanoantennas. However, these signals do not carry particular phase information but can contribute to the background signal. Furthermore, the speckle noise in the holographic images can be explained by the high sensitivity of the nanoantenna size. The limited fabrication accuracy can result in phase noise, which causes deviations from the calculated hologram shown in Figure 3B. The circular background shape originates from the *k*-space limit of the microscope objective in our imaging system.

*Spectral characterization*

The spectral analysis of the SHG and THG signal is measured in the same configuration as the images (Fig. 3A), except that the camera is replaced by a spectrograph. The incident laser power is tuned from 2.7 kW to 13.3 kW at the operation-wavelength of 1290 nm with a full width at half maximum (FWHM) of 17 nm and the corresponding spectra are recorded. Figure 4A shows the spectra of the measured nonlinear signals from the metasurface hologram. The SHG signal shows a peak located at a wavelength of 645 nm with a FWHM of 7 nm while the THG signal is located at 430 nm with an FWHM of 5 nm. As expected from the different power dependencies of both harmonic processes, the third-order process scales stronger than the second-order process. From 550 nm to longer wavelengths, we find a broad background signal, which can be explained by the photoluminescence of gold nanostructures [38]. Further, we integrated the spectral intensities, depending on the incident laser power (Fig. 4B). The double logarithmic plotted intensities over incident laser power follow a quadratic and a cubic behavior of the SHG and THG process, respectively [39]. The red and blue dashed lines in the graph correspond to the fit functions:

$$I_{SHG}^{fit} = 2.7 \cdot 10^{-9} \frac{1}{mW^2} P^2$$

$$I_{THG}^{fit} = 4.4 \cdot 10^{-13} \frac{1}{mW^3} P^3$$

where *P* is the average power of the laser beam. From a spectral point of view, the nonlinear signal intensities are equal for a power of 6.1 kW at 1290 nm incident wavelength. This value is slightly lower than obtained for the imaging system with the CMOS camera (Fig. 3), where we estimated the equal intensity point at around 10.7 kW. The deviations result from the different spectral responses and the distribution of the SHG and THG signals to larger area sizes on the detector, which lowers the signal-to-noise ratio and increases the error.

**Conclusion**

In conclusion, we show a two-colored nonlinear hologram based on the second and third harmonic generation of near-infrared light. The plasmonic metasurface hologram consists of an equivalent amount of C2 and C3 rotational symmetric gold nanoantennas with subwavelength thickness. Based on the nonlinear Pancharatnam-Berry phase, we encoded a dual-color phase-only hologram in the orientation of the C2 and C3 nanoantennas to generate a visible holographic image. We show that a balanced signal ratio of both nonlinear signals, SHG and THG, can form a bi-colored image. Further, the nonlinear response can be modulated depending on the structure's geometry, as well as on the incident wavelength and laser power. Thus, it is in general possible to engineer multi-harmonic nonlinear optical devices with adapted strength of the involved harmonic processes and simultaneously encode spatial phase information. Our approach demonstrates an alternative way to realize colored holograms for illumination with monochromatic light. The concept can be further extended to obtain multi-harmonic optical devices that can show different beam shaping functionalities for different nonlinear processes.

**Acknowledgments**

This project received funding from the Deutsche Forschungsgemeinschaft (DFG, German Research Foundation) – SFB-Geschäftszeichen TRR142/2-2020 – Projektnummer 231447078 – Teilprojekt A08. The authors acknowledge the funding provided by the National Key R&D Program of China (No. 2017YFB1002900) and the European Research Council (ERC) under the European Union's Horizon 2020 research and innovation program (grant agreement No. 724306). We also acknowledge the NSFC-DFG joint program (DFG No. ZE953/11-1, NSFC No. 61861136010). L.H. acknowledges the support from the Beijing Outstanding Young Scientist Program (BJJWZYJH01201910007022) and the National Natural Science Foundation of China (No. 61775019) program.

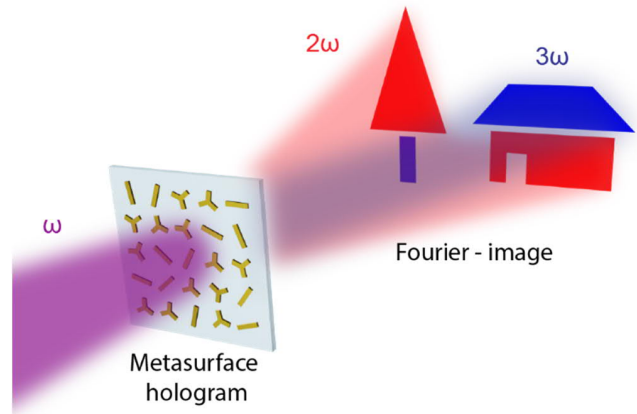

**Figure 1. Plasmonic Metasurface for nonlinear holographic image encryption.** The plasmonic meta-atoms exhibit different rotation symmetries and support different harmonic generations. Based on the Pancharatnam-Berry phase, a holographic image can be encoded, while the position of the metaatoms can be randomly distributed in a square lattice, but the rotation of each antenna is significant for the image encryption. The schematic illustration shows a nonlinear plasmonic metasurface consisting of C2 and C3 gold nanoantennas, that can carry holographic information, for instance, a colored tree and house. If the metasurface is illuminated at its resonance frequency ω, the image is generated at frequencies 2ω and 3ω.

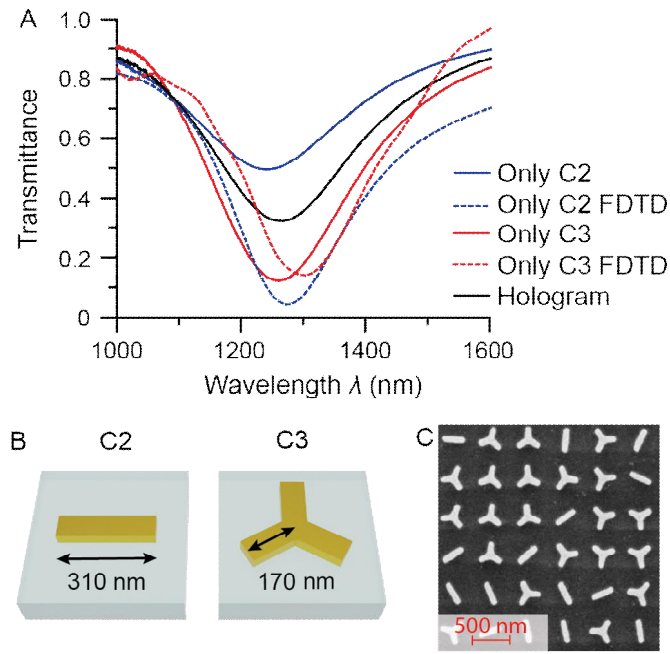

**Figure 2. Linear transmittance and metasurface design.** A) Transmittance of the nonlinear color hologram (black) for unpolarized incident light, illustrating the plasmon resonance in the near-infrared. The transmittance of single C2 (blue) and single C3 (red) structure metasurfaces confirm the resonance frequency of the combined hologram. The blue and red dashed lines show the simulated transmittance of a C2 only and C3 only metasurface. B) The unit cell size of the metasurface is $500 \times 500$ nm$^2$. The arm length of the C2 and C3 structures is tailored to obtain an equal resonance. All structures are fabricated based on electron beam lithography and a lift-off process. The particles consist of 30 nm gold. Three-fold (C3) and two-fold (C2) rotation symmetric gold nanoparticles are used for second- and third-harmonic generation, respectively. C) SEM image of the nonlinear color hologram with randomized antenna placement in a square lattice.

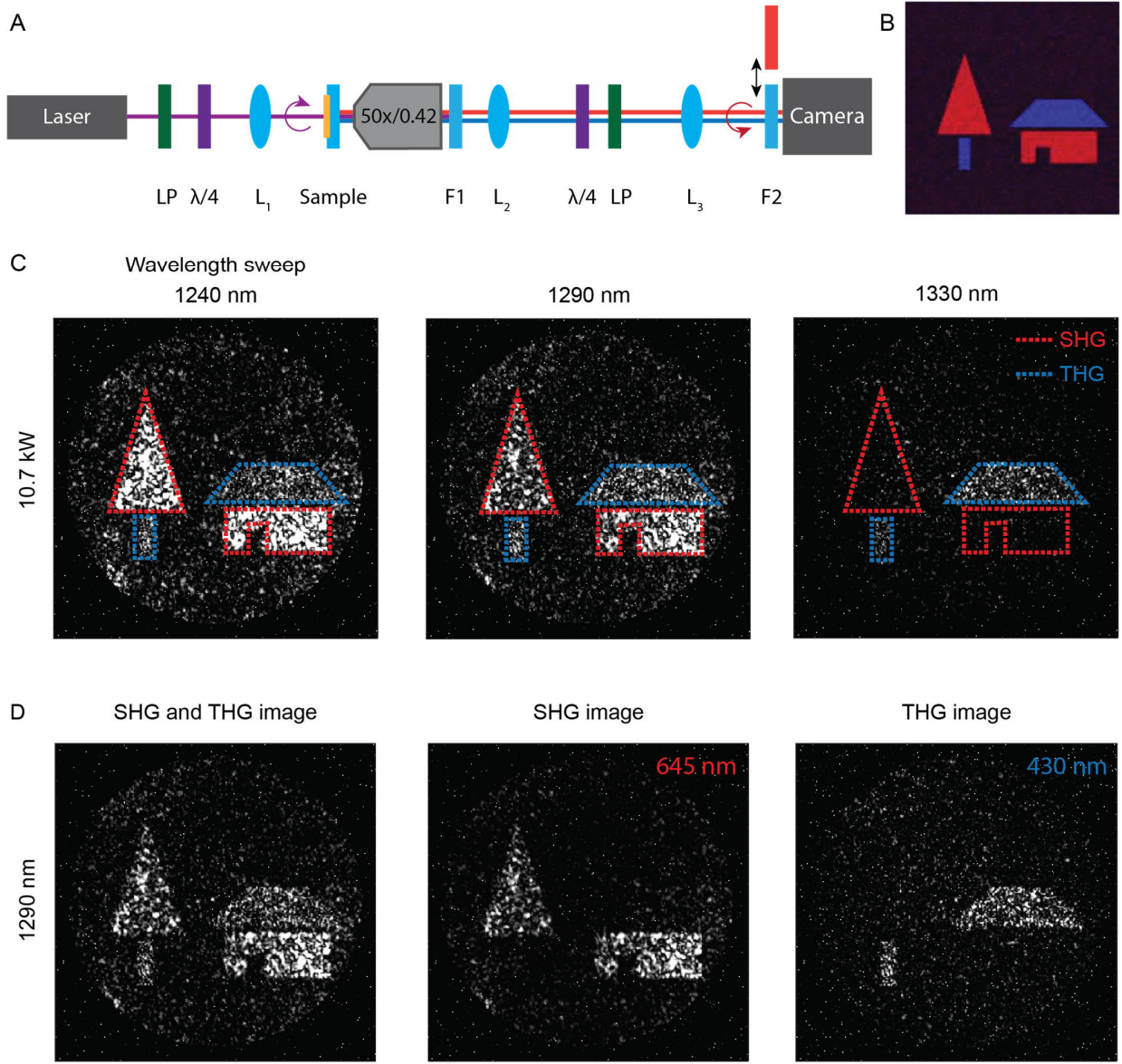

**Figure 3. Two-color nonlinear meta-hologram.** A) Laser light is circularly polarized using a linear polarizer and a quarter-wave plate. The light is focused on the metasurface and the transmitted light is collected with a microscope objective (50x/NA = 0.42). The filter $F_1$ blocks the fundamental wave, while the color filters $F_2$ are used to choose different wavelength channels. For imaging, we use two additional lenses $L_{2,3}$ with a focal length of 100 mm. An additional polarizer unit is used to image the cross-polarization channel on the camera. B) Colored simulated reconstruction of the hologram. C) Spectral behavior of the Fourier-image. By changing the incident wavelength of the laser, the ratio between SHG and THG signal can be changed. The red and blue dashed lines indicate areas of weak SHG and THG signals, respectively. D) Fourier-image of the nonlinear two-color hologram (SHG and THG). The wavelength of the circularly polarized input beam is 1290 nm. We use different filters to image the SHG (645 nm) and THG (430 nm) signals separately.

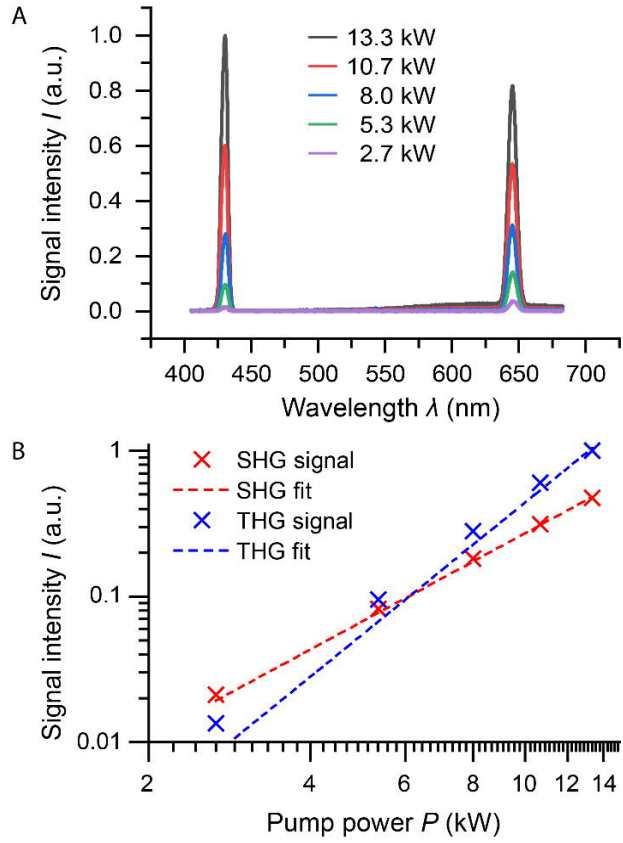

**Figure 4. Nonlinear spectral response of the nonlinear hologram.** A) Nonlinear response of the holographic metasurface for a pump wavelength of 1290 nm and peak powers between 2.7 kW and 13.3 kW. The THG and SHG signal is located at 430 nm and 645 nm, respectively. B) Integrated spectral powers of the SHG and THG signals, respectively. The red and blue dashed lines indicate the quadratic and cubic fit function, respectively. We observe a crossing point between the SHG and THG signals at about 6.1 kW.